\documentclass[twocolumn,showpacs,prl,superscriptaddress,a4paper,floatfix]{revtex4}
\usepackage{graphicx}
\usepackage{amsmath}
\usepackage{graphicx}
\usepackage{dcolumn}
\usepackage{bm}

\begin{document}

\title{Importance of bath dynamics for decoherence in spin systems}
\author{Shengjun Yuan}
\affiliation{Department of Applied Physics, Zernike Institute for Advanced Materials,
University of Groningen, Nijenborgh 4, NL-9747 AG Groningen, The Netherlands}
\author{Mikhail I. Katsnelson}
\affiliation{Institute of Molecules and Materials, Radboud University of Nijmegen, 6525
ED Nijmegen, The Netherlands}
\author{Hans De Raedt}
\affiliation{Department of Applied Physics, Zernike Institute for Advanced Materials,
University of Groningen, Nijenborgh 4, NL-9747 AG Groningen, The Netherlands}
\date{\today }

\begin{abstract}
We study the decoherence of two coupled spins that interact with a
chaotic spin-bath environment. It is shown that connectivity of
spins in the bath is of crucial importance for the decoherence of
the central system. The previously found phenomenon of two-step
decoherence (Phys. Rev. Lett. {\bf 90}, 210401 (2003)) turns out
to be typical for the bath with a slow enough dynamics or no
dynamics.
For a generic random system with chaotic dynamics a conventional exponential
relaxation to the pointer states takes place. Our results confirm
a conjecture of Paz and Zurek (Phys. Rev. Lett. {\bf 82}, 5181
(1999)) that for weak enough interactions the pointer states are
eigenstates of the central system.
\end{abstract}

\pacs{03.67.Mn, 05.45.Pq, 75.10.Nr }
\maketitle




It is commonly accepted that decoherence by nuclear spins is
the main obstacle for realization of quantum computations in
magnetic systems; see, e.g., discussions of specific silicon
\cite{kane} and carbon \cite{loss} based quantum computers.
Therefore, understanding the decoherence in quantum spin systems is a
subject of numerous works (for a review, see Refs.~\onlinecite{stamp,Slava2006}).
The issue seems to be very complicated and despite many efforts,
even some basic questions about character of the decoherence process are unsolved yet.
Most of the problems cannot be solved analytically, in particular if
there is more than one spin in the central system, but we
can use the computer to simulate the dynamics and find useful
information.

An unusual two-step decoherence was reported in
Ref.~\onlinecite{ourPRL}. This is an important phenomenon since it
implies that, generally speaking, the observation of the Rabi
oscillations does not guarantee access to sectors
of the Hilbert space that may be essential for efficient quantum computation.
Its origin is still poorly understood; it was
described analytically in a framework of an exactly solvable model
of noninteracting spins in the bath \cite{Melikidze2004} but it is
not clear how sensitive it is to the details of spin-spin
interactions. In the real world, the environment has its
own dynamics, which could be much slower or comparable to the
central dynamics. First attempts to investigate numerically the
effects of the environment dynamics \cite{ourPRE} did not lead to
definite conclusions.

The behavior of an open quantum system crucially depends on
the ratio of typical energy differences of the central system
$\delta E_c$ and the energy $E_{ce}$ which characterizes
the interaction of the central system with the environment.
The case $\delta E_c \ll E_{ce}$ has been studied
extensively in relation to the ``Schr\"{o}dinger cat'' problem and the physics is quite clear~ \cite{zeh,zurek}:
As a result of time evolution, the central system
passes to one of the ``pointer states''~\cite{zurek} which, in this case, are the
eigenstates of the interaction Hamiltonian. The opposite case,
$\delta E_c \gg E_{ce}$ is less well understood. There is a
conjecture that in this case the pointer states should be
eigenstates of the Hamiltonian of the central system, but this
is proven only for a very simple model~\cite{paz}. On the
other hand, this case is of primary interest if, say, the central
system consists of electron spins whereas the environment are
nuclear spins (e.g., if one considers the possibility of quantum
computation using molecular magnets \cite{m1,m2}).

In fact, as we will show, the selection of an eigenstate as the
pointer state is also determined by the state and the dynamics of
the environment. Elsewhere~\cite{JETPLett,Yuan2007}, we have already shown
that if the environment is a spin glass 
and initially in the ground state, then independent
of the initial state of the central system, the central system
relaxes to a state that is very close to its ground state:
The ground state is selected as the point state of the central system.
In this Letter, we consider a realistic model of
decoherence of a system of two spins by an environment of nuclear spins
at elevated temperatures.
We will demonstrate that the decoherence of the central system
depends in a significant, nonintuitive manner on the details
of the dynamics of the environment.

We consider a generic quantum spin model described by the
Hamiltonian $H =H_{c}+H_{ce}+H_{e}$ where
\begin{align}
H_{c}& =-J\mathbf{S}_{1}\cdot \mathbf{S}_{2},  \notag \\
H_{e}& =-\sum_{i=1}^{N-1}\sum_{j=i+1}^{N}\sum_{\alpha }\Omega
_{i,j}^{(\alpha )}I_{i}^{\alpha }I_{j}^{\alpha },  \notag \\
\noalign{and}
H_{ce}& =-\sum_{i=1}^{2}\sum_{j=1}^{N}\sum_{\alpha }\Delta _{i,j}^{(\alpha)}S_{i}^{\alpha }I_{j}^{\alpha },
\label{HAM}
\end{align}
are the Hamiltonians of the central system, the environment,
and the interaction between the central system and the environment, respectively.
In Eq.~(\ref{HAM}), the exchange integrals $J$ and $\Omega _{i,j}^{(\alpha )}$ determine
the strength of the interaction between spins
$\mathbf{S}_{n}=(S_{n}^{x},S_{n}^{y},S_{n}^{z})$ of the central system $H_{c}$, and
the spins $\mathbf{I}_{n}=( I_{n}^{x},I_{n}^{y},I_{n}^{z} ) $ in
the environment $H_{e}$, respectively. The exchange integrals $\Delta_{i,j}^{(\alpha )}$
control the interaction $H_{ce}$ of the central system
with its environment. In Eq.~(\ref{HAM}), the sum over $\alpha $ runs over
the $x$, $y$ and $z$ components of spin-$1/2$ operators $\mathbf{S}$ and $\mathbf{I}$.

In the sequel, we will use the term ``Heisenberg-like'' $H_{e}$ 
to indicate that each $\Omega _{i,j}^{(\alpha )}$ 
is a  uniform random number in the range $[-\Omega |J|,\Omega |J|]$, 
$\Omega $ being a free parameter.
We will consider two different kinds of $H_{ce}$, namely
rotational invariant Heisenberg interactions $\Delta_{i,j}^{(\alpha )}\equiv \Delta $
and ``Ising-like'' interactions for which $\Delta _{i,j}^{(x)}=\Delta _{i,j}^{(y)}=0$ and
$\Delta _{i,j}^{(z)}$ are dichotomic random variables, taking the values $\pm \Delta $.
Obviously, if $H_{ce}$ is ``Ising like'', the
total magnetization of the central system ($M=S^z_1+S^z_2$) is
a conserved quantity.

%
As we demonstrate in this Letter, the connectivity of the spins
in the environment affects the decoherence in a nontrivial manner.
We characterize this connectivity by the number $K$, the number
of environment spins with which a spin in the environment interacts.
If $K=0$, each spin in the environment interacts with the central system only.
If $K=2$, the structure of the environment is assumed to be that of a ring, that is
each spin in the environment interacts with two other spins only.
Likewise, $K=4$ and $K=6$ correspond environments in which the spins
are placed on a square or triangular lattice, respectively and interact with
nearest-neighbors only.
If $K=N$, each spin in the environment interacts with all the other spins
in the environment and, to give this case a name, we will refer
to this case as ``spin glass''.

If the Hamiltonian of the central system $H_{c}$ is a
perturbation, relative to the interaction Hamiltonian $H_{ce}$,
the pointer states are eigenstates of $H_{ce}$~\cite{zurek}. In
the opposite case, that is the regime $|\Delta| \ll |J|$ that we
explore in this paper, the pointer states are supposed to be
eigenstates of $H_{c}$~\cite{paz}. The latter are given by $\vert
1\rangle \equiv \vert T_1\rangle =\vert \uparrow \uparrow \rangle
$, $\vert 2\rangle \equiv \vert S\rangle =(\vert \uparrow
\downarrow \rangle -\vert \downarrow \uparrow \rangle)/\sqrt{2}$,
$\vert 3\rangle \equiv \vert T_0\rangle =(\vert \uparrow
\downarrow \rangle +\vert \downarrow \uparrow \rangle)/\sqrt{2}$,
and $\vert 4\rangle \equiv \vert T_{-1}\rangle =\vert \downarrow
\downarrow \rangle $, satisfying $H_{c}\vert S\rangle =
(3J/4)\vert S\rangle$ and $H_{c}\vert T_{i}\rangle =(-J/4)\vert
T_{i}\rangle$ for $i=-1,0,1$. To check this conjecture is one of
the main aims of our simulations.

The simulation procedure is as follows.
We generate a random superposition $\left\vert \phi\right\rangle $
of all the basis states of the environment.
This state corresponds to the equilibrium
density matrix of the environment at infinite temperature.
The spin-up -- spin-down state ($|\uparrow \downarrow \rangle $) is taken
as the initial state of the central system. Thus, the initial state of the whole system reads
$\left\vert \Psi (t=0)\right\rangle \rangle =\left\vert\uparrow \downarrow \right\rangle \left\vert \phi \right\rangle $
and is a product state of the state of the central system and the random state of the
environment which, in general is a (very complicated) linear combination of
the $2^{N}$ basis states of the environment.
In our simulations we take $N=16$ which, from earlier work~\cite{JETPLett,Yuan2007}, we
know is sufficiently large for the environment to behave as a ``large'' system.
For a given, fixed set of model parameters, the time evolution of
the whole system is obtained by solving the time-dependent Schr\"{o}dinger equation 
for the many-body wave function $|\Psi (t)\rangle $, describing
the central system plus the environment. The numerical method that
we use is described in Ref.~\onlinecite{method}. It conserves the
energy of the whole system to machine precision. We monitor the
effects of the decoherence by plotting the time dependence of the
matrix elements of the reduced density matrix of the central
system. As explained earlier, in the regime of interest $|\Delta|
\ll |J|$, the pointer states are the eigenstates of the central
systems, hence we compute the matrix elements of the density
matrix in the basis of eigenvectors of the central system.

\begin{figure}[t]
\begin{center}
\includegraphics[width=9cm]{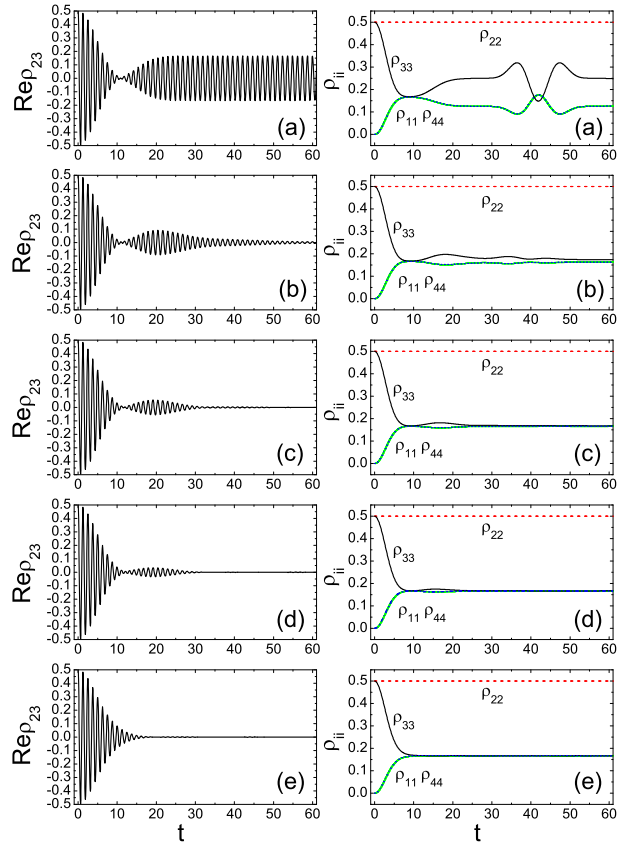}
\end{center}
\caption{(Color online) The time evolution
of
the real part of the off-diagonal element $\rho _{23}$
(left panel)
and
the diagonal elements $\rho _{11},\ldots, \rho _{44}$
(right panel)
of the reduced density matrix of a central system (with $J=-5$),
interacting with a Heisenberg-like environment $H_{e}$ (with $\Omega =0.15$)
via an isotropic Heisenberg Hamiltonian $H_{ce}$ (with $\Delta =-0.075$ )
for different connectivity numbers $K$ of the spins in the environment: 
(a) $K=0$; (b) $K=2$; (c) $K=4$; (d) $K=6$; (e) $K=N$.
The absolute values of the off-diagonal elements $\rho _{12}$,
$\rho _{13}$,
$\rho _{14}$,
$\rho _{24}$,
and
$\rho _{34}$ are less than 0.002 (results not shown).
}
\label{fig1}
\end{figure}

In Fig.~\ref{fig1}, we show the time evolution of the
elements of the reduced density matrix of the central system
for different connectivity numbers $K$ with isotropic Heisenberg
interactions between the central system and the spins in the environment.
We conclude that:
\begin{enumerate}
\item{In agreement with earlier work~\cite{ourPRL,Melikidze2004},
we find that in the absence of interactions between the
environment spins ($K=0$) and after the initial fast decay, the
central system exhibits long-time oscillations (see
Fig.~\ref{fig1}(a)(left)). In this case and in the limit of a
large environment, we have~\cite{Melikidze2004}
\begin{equation}
\hbox{Re }\rho _{23}\left( t\right) =\left[ \frac{1}{6}+
\frac{(1-bt^{2}) e^{-ct^{2}}}{3}\right] \cos \omega t,
\label{melik}
\end{equation}
where $b=N\Delta ^{2}/4$, $c=b/2$ and $\omega=J-\Delta$.
Equation~(\ref{melik}) clearly shows the two-step process,
that is, after the initial Gaussian decay of the amplitude of the oscillations,
the oscillations revive and their amplitude levels of~\cite{Melikidze2004}.
Notice that because of conservation laws, this behavior does not change if the environment
is described by an isotropic Heisenberg Hamiltonian
($\Omega _{i,j}^{(\alpha )}\equiv \Omega $ for all $\alpha$, $i$ and $j$), whatever the value
of $K$.
If, as in Ref.~\cite{ourPRL}, we take $\Delta_{i,j}^{(x)}=\Delta _{i,j}^{(y)}=\Delta _{i,j}^{(z)}\in \left[ 0,\Delta %
\right] $ random instead of the same,
the amplitude of the long-living oscillations is no longer constant
but decays very slowly~\cite{ourPRL}.
}
\item{The presence of Heisenberg-like (non-isotropic, random) interactions between the spins
of the environment leads to a reduction and a decay of the amplitude of the long-living oscillations
(see Fig.~\ref{fig1}(b-e)(left)).
The larger the connectivity number $K$, the faster is the decay of the amplitude.
When $K$ reaches its maximum $K=N$ (spin glass), the second step in the decoherence
process is no longer separated from the initial decay.
In fact, it seems as if it has merged with the final stage of the first step
(see Fig.~\ref{fig1}(e)(left)).
For $K=N$, the time evolution of $\rho _{23}\left( t\right) $ can be fitted
well by the function
\begin{equation}
\hbox{Re }\rho _{23}\left( t\right) =\left[ \frac{e^{-a't}}{6}+
\frac{(1-b't^{2}) e^{-c't^{2}}}{3}\right] \cos \omega' t,
\end{equation}
with $a'=0.13403$, $b'=0.00659$, $c'=0.01085$ and $\omega'=5.01037$.
For comparison, the values that enter Eq.~(\ref{melik}) are
$b=0.0225$, $c=0.01125$ and $\omega=4.925$.
It is of interest to note that if the dynamics of the environment
is sufficiently slow ($\Omega \approx0.01$ in Ref.~\cite{ourPRL}),
this dynamics apparently does not affect the decoherence of the central spins~\cite{ourPRL}.
Thus, the effectiveness of the decoherence process not only depends on $K$ but also on the
details of the interactions within the environment.
}
\item{
According to the general picture of decoherence~\cite{zurek},
for an environment with nontrivial internal dynamics
that initially is in a random superposition of all its eigenstates,
we expect that the central system will evolve to a
stable mixture of its eigenstates.
In other words, the decoherence will cause all the off-diagonal elements
of the reduced density matrix to vanish with time.
Moreover, the weight of the degenerate eigenstates $\vert T_{0}\rangle$,
$\vert T_{-1}\rangle$, and $\vert T_{1}\rangle$ in
this mixed state are expected to be the same.
As shown in Fig.~\ref{fig1}(b-e)(right), our simulations
confirm that this picture is correct in all respects.
Furthermore, the results depicted in Fig.~\ref{fig1}(b-e)(right)
demonstrate that the connectivity number $K$ has no
effect on the value of $\rho _{11}$, $\rho _{22}$,
$\rho _{33}$ and $\rho _{44}$ for long times.
}
\end{enumerate}

\begin{figure}[t]
\begin{center}
\includegraphics[width=9cm]{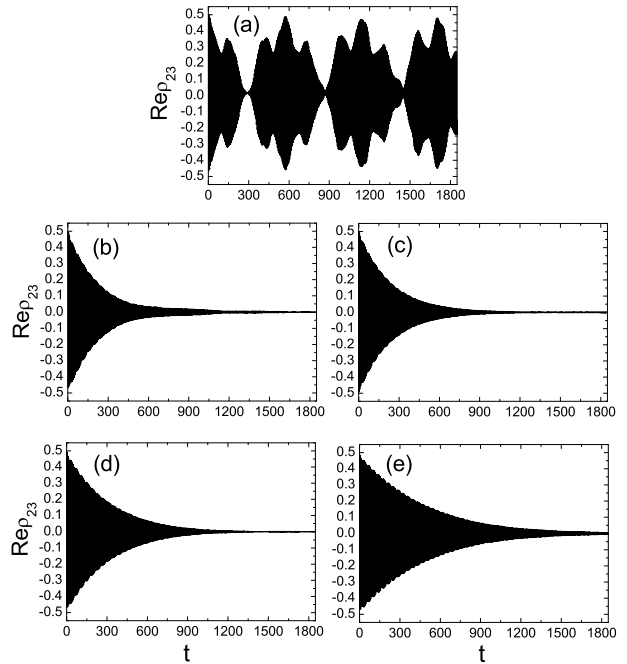}
\end{center}
\caption{The time evolution of the real part of the off-diagonal element $\rho _{23}$
of the reduced density matrix of a central system (with $J=-1$),
interacting with a Heisenberg-like environment $H_{e}$ (with $\Omega =0.15$)
via an Ising-like Hamiltonian $H_{ce}$ (with $\Delta =0.075$ )
for different connectivity numbers $K$ of the spins in the environment: 
(a) $K=0$; (b) $K=2$; (c) $K=4$; (d) $K=6$; (e) $K=N$.
}
\label{fig4}
\end{figure}

\begin{table}[t]
\caption{Frequency of oscillation $\omega''$ and decoherence time $\tau$
for different connectivity $K$.
In case 1 and case 2, for each $K$, the values of $\Omega_{i,j}^{(\alpha)}$ are the same
but the values of $\Delta_{i,j}^{(\alpha)}$ are different.
In case 2 and case 3, for each $K$, the values of
$\Delta_{i,j}^{(\alpha)}$
are the same
but the values of
$\Omega_{i,j}^{(\alpha)}$
are different.
}
\label{table2}
\begin{center}
\begin{ruledtabular}
\begin{tabular}{cccccc}
Case  & & $K=2$ & $K=4$ & $K=6$ & $K=N$\\ \hline
\noalign{\smallskip}\hline\noalign{\smallskip}
 1 & $\omega''$ & $1.015$ & $1.015$ & $1.016$ & $1.017$\\
  & $\tau''$ & $212.9$ & $235.3$ & $302.9$ &  $447.4$\\
\noalign{\smallskip}\hline\noalign{\smallskip}
2 & $\omega''$ & $1.017$ & $1.021$ & $1.022$ & $1.027$\\
  & $\tau''$ & $93.07$ & $105.0$ & $110.4$ & $182.1$ \\
\noalign{\smallskip}\hline\noalign{\smallskip}
3 & $\omega''$ & $1.021$ & $1.022$ & $1.022$ & $1.027$\\
 & $\tau''$ & $106.3$ & $110.0$ & $111.7$ & $173.2$ \\
\end{tabular}
\end{ruledtabular}
\end{center}
\end{table}
\qquad \qquad\

Next, we change the interactions between central system and the spins in the environment
from Heisenberg-like to Ising-like but keep the interactions between different spins in the environment
Heisenberg-like.
In fact, in our simulations, the Hamiltonian $H_{e}$ is the same for both cases.
As explained earlier, in the Ising-like case, the total magnetization
of the central system is conserved during the time evolution.
Thus, as the initial state of the central system is $(\vert S\rangle+\vert T_0\rangle)/\sqrt{2}$,
at any time $t$ the state of the whole system can be written as
\begin{equation}
|\Psi (t)\rangle =|2\rangle |\phi_2(t)\rangle +|3\rangle |\phi_3(t)\rangle ,  \label{Phia}
\end{equation}
where $|\phi _2(t)\rangle $ and $|\phi _3(t)\rangle$ denote the states of the environment.
In other words, only $\rho_{22}(t)$, $\rho_{23}(t)$, $\rho_{32}(t)$, and
$\rho_{33}(t)$ can be nonzero.

On general grounds, we may expect that the presence of an additional
conservation law slows down the decoherence and indeed, as
shown in Fig.~\ref{fig4}, this is the case.
Note that the results of Fig.~\ref{fig4} have been
obtained for $J=-1$ instead of for $J=-5$, the value used
to compute the results shown in Fig.~\ref{fig1}
(the latter value was chosen to facilitate the comparison with
the results of Ref.~\cite{ourPRL}),
but this factor of five in the value of $J$ cannot account
for the large difference in the observed decoherence times.

From Fig.~\ref{fig4}, we conclude that
\begin{enumerate}
\item{We never observe the two-step process that
we find in the case of Heisenberg-like $H_{ce}$.
For $K=0$ (see Fig.~\ref{fig4}(a)), there is no decoherence.}
\item{For $K>0$, $\hbox{Re }\rho_{23}(t)$ vanishes with time, in agreement
with the general picture of decoherence~\cite{zurek}.
However, quite unexpectedly, the rate of decoherence {\sl increases}
with $K$, in contrast to the case of Heisenberg $H_{ce}$
in which the rate of decoherence decreases with $K$.
The data presented in Figs.~\ref{fig4}(b-d) can all be fitted
very well by
\begin{equation}
\hbox{Re }\rho _{23}\left( t\right) =\frac{1}{2}e^{-t/\tau''}\cos \omega'' t,
\end{equation}
where $\omega'' \approx |J|$ and the values of $\tau''$ depend on $K$.
In Table \ref{table2}, we give some typical results for these
parameters. Cases 1 and 2 illustrate that (random) changes in $H_{ce}$ may affect
the value of the decoherence time $\tau''$ significantly but the general trend,
the increase of $\tau''$ with $K$ seems generic.
}
\end{enumerate}

In conclusion, we have shown that
(1) the pure quantum state of the central spin system evolves into the classical, mixed state,
and (2) if the interaction between the central system and environment is much smaller
than the coupling between the spins in the central system, the pointer states are the eigenstates
of the central system.
Both these observations are in concert with the general picture of decoherence~\cite{zurek}.

Furthermore, we have demonstrated that, in the case that the
environment is a spin system, the details of this spin system are
important for the decoherence of the central system. In
particular, we have shown that (1) changing the internal dynamics
of the environment may change the qualitative features of the
decoherence of the central spin system, and that (2) the
dependence of the decoherence time of the central spin system on
the connectivity of the interactions between spins of the
environment is counterintuitive.

\section*{Acknowledgements}

M.I.K. acknowledges a support by the Stichting
Fundamenteel Onderzoek der Materie (FOM).

\end{document}